\documentclass[aps,superscriptaddress,twocolumn]{revtex4}
\usepackage{amsmath,graphicx,cancel,color}
\usepackage{color}
\usepackage{amsthm}
\usepackage{amssymb}
\usepackage{multirow}
\begin{document}
\title{Assessing student reasoning in upper-division electricity and magnetism\\
at Oregon State University}

\author{Justyna P. Zwolak}
\email[]{j.p.zwolak@gmail.com}
\affiliation{Department of Physics, Oregon State University, Corvallis, Oregon 97331, USA}
\affiliation{STEM Transformation Institute, Florida International University, Miami, Florida 33199}
\author{Corinne A. Manogue}
\affiliation{Department of Physics, Oregon State University, Corvallis, Oregon 97331, USA}

\date{\today}


\begin{abstract}
Standardized assessment tests that allow researchers to compare the performance of students under various curricula are highly desirable. There are several research-based conceptual tests that serve as instruments to assess and identify students' difficulties in lower-division courses. At the upper-division level assessing students' difficulties is a more challenging task. Although several research groups are currently working on such tests, their reliability and validity are still under investigation. We analyze the results of the Colorado Upper-Division Electrostatics diagnostic from Oregon State University and compare it with data from University of Colorado. In particular, we show potential shortcomings in the Oregon State University curriculum regarding separation of variables and boundary conditions, as well as uncover weaknesses of the rubric to the free response version of the diagnostic. We also demonstrate how the diagnostic can be used to obtain information about student learning during a gap in instruction. Our work complements and extends the previous findings from the University of Colorado by highlighting important differences in student learning that may be related to the curriculum, illuminating difficulties with the rubric for certain problems and verifying decay in post-test results over time.
\end{abstract}


\maketitle

\section{Background and motivation}

Designing standardized assessment tests that allow researchers to compare the performance of students taught according to various curricula is one of the primary tasks of education research. Such comparisons provide information about the relative effectiveness of different curricula and, as a result, can improve methods of teaching, learning trajectories and, ultimately, student learning. Appropriately designed diagnostics not only reveal common student difficulties but can also help to determine to what extent students understood the content. 

As of the present day, there are several research-based conceptual tests that serve as instruments to assess and identify students' difficulties in lower-division courses (e.g., the Force Concept Inventory \cite{Hestenes92-FCI}, the Conceptual Survey of E\&M \cite{Maloney01-CSEM} and the Brief Electricity and Magnetism Survey \cite{Ding06-BEMA}). Data from these tests help to determine, among other things, where students lack a conceptual understanding of the material and help to correlate this with various methods of teaching. It also allows teachers and researchers to find out if these difficulties are widespread. 

Assessing students' difficulties is more intricate at the upper-division level, in part due to the increased complexity of the content. It is harder to design a rubric that will include all possible approaches to a problem. At the same time, a rigorous rubric is necessary to assure consistency in grading between different institutions. Several research groups are currently working on such diagnostic tests, e.g., the Colorado Upper-Division Electrostatics \cite{Chasteen09-CUE1,Chasteen12-CUE,Pepper12-SDM}, the Colorado UppeR-division ElectrodyNamics Test \cite{Baily12-AUE}, the Quantum Mechanics Assessment Tool \cite{Goldhaber09-TQM} and the Survey of Quantum Mechanics Concepts \cite{Singh05-AQM}. These upper-division assessments are relatively new and have only been employed at a few institutions. Thus, their validity and robustness when used at institutions outside their place of origin is still an active area of investigation. 

In the Paradigms in Physics program at Oregon State University (OSU), we instituted a radical reform of all the upper-division physics courses that led to extensive reordering of the content. Thus, our program represents an important test case to examine the versatility of this new assessment tool. 

In this paper, we present our findings from the analysis of data collected at OSU using one of the measures developed at the University of Colorado at Boulder (CU) for upper-division electricity and magnetism I (E\&M I) -- the Colorado Upper-Division Electrostatics diagnostic (CUE). We address three main questions: 1) What does the CUE tell us about students' learning at OSU? In particular, we discuss how the scores compare between institutions and what differences between curricula the CUE can reveal. 2) What does the data from two different institutions tell us about the CUE? We discuss how the rubric reflects students' knowledge and the issues uncovered by the multiple choice version of the CUE. 3) What information can be obtained from the midtest -- an additional CUE test that was introduced at OSU?

\begin{table*}[t]
\caption{Standard schedule of Paradigms (junior year courses) and Capstones (senior year courses). E\&M-related courses, during which the CUE is being administered, are highlighted in bold. Beginning in academic year 2011/12 the Mathematical Methods and Classical Mechanics courses switched places, with the former now coming in the Spring and the latter in the Fall.}
\renewcommand{\arraystretch}{1.2}
\begin{tabular}{c c c || c c}
\cline{1-5} 
\multicolumn{3}{c||}{Junior Courses} & \multicolumn{2}{c}{Senior Courses} \\ 
Fall & Winter & Spring & Fall & Winter \\ \cline{1-5}
\rule{0pt}{25pt} \parbox[c]{3cm}{\centering \textbf{Symmetries}\\ \textbf{Vector Fields}\\ Oscillations} &\parbox[c]{3cm}{\centering Preface\\Spins\\1-D Waves\\Central Forces\vspace{3pt}} & \parbox[c]{3cm}{\centering Energy and Entropy\\Periodic Systems\\Reference Frames\\Classical Mechanics\vspace{2pt}} & \parbox[c]{3cm}{\centering Mathematical Methods\\\textbf{Electromagnetism}} & \parbox[c]{3cm}{\centering Quantum Mechanics\\ Statistical Physics\\ Physical Optics}\\ 
\cline{1-5}
\end{tabular}
\label{tab:course_desc}
\end{table*}

The paper is organized as follows: We start with a detailed description of the Paradigms curriculum in Sec.~\ref{sec:curr_at_OSU} and the methodology in Sec.~\ref{sec:method}. Then we move to a discussion of the overall findings from OSU. In Sec.~\ref{sec:what_CUE_tells_us}, we present the general analysis of OSU students' performance and discuss difficulties revealed using the CUE. In Sec.~\ref{sec:what_curriculum_tells_us}, we look more closely at specific questions from the CUE, uncovering problems with the grading rubric. We discuss differences between the free response and multiple choice versions of the CUE and possible reasons for why students' answers might not fit the classification of the rubric in its current form. Finally, in Sec.~\ref{sec:gap_in_instruction} we examine the long-term learning of students at OSU using the newly introduced CUE midtest. We conclude with a discussion of future research directions in Sec.~\ref{sec:summary}.

\section{Curriculum at OSU}\label{sec:curr_at_OSU}

OSU's middle- and upper-division curriculum was extensively reorganized in 1997 compared to traditionally taught courses. This led to a substantial reordering of the content \cite{Manogue01-TUC}. In traditional curricula, courses focus on a particular subfield of physics (e.g., classical mechanics, electricity and magnetism, quantum mechanics). A first one-semester E\&M~I course (15-16 weeks) at a research university covers approximately the first six chapters of the standard text ``Introduction to Electrodynamics" by David J. Griffiths \cite{Griffiths-EM}, i.e.,  a review of the vector calculus necessary for a mathematical approach to electricity and magnetism, as well as electrostatics and magnetostatics both in a vacuum and in matter.  A second  semester course on electrodynamics (E\&M II) would typically cover most of the remaining chapters of Griffiths.

At OSU, junior-level courses -- called Paradigms -- revolve around concepts underlying the physics subfields (e.g., energy, symmetry, forces, wave motion; see Table~\ref{tab:course_desc} for a course schedule). Therefore, the content is arranged differently and certain topics are emphasized more than in traditional courses. For instance, in E\&M-related Paradigms (``Symmetries'' and ``Vector Fields'') more time is spent on direct integration and curvilinear coordinates, and less time on separation of variables. There is also variation in the sequence --  potentials are discussed before electric fields and magnetostatics in a vacuum before electrostatics in matter. We integrate the mathematical methods with the physics content, including a strong emphasis on off-axis (i.e., non-symmetric) problems and power series approximations. 

The first two Paradigms cover electro- and magnetostatics in a vacuum, approximately the material covered in Griffiths Chapters 1, 2 and 5. The gravitational analogue of electrostatics is covered at the same time as electrostatics rather than in a classical mechanics course and the method of separation of variables is discussed as part of the quantum mechanics Paradigms (``1-D Waves'' and ``Central Forces''). We also use a large variety of active engagement strategies, such as individual small white board questions, small group problem-solving,  kinesthetic activities, computer visualizations, simulations and animations \cite{OSU-activ}.
 
The Paradigms courses, taken in the junior year, are followed by Capstones courses, which have a more traditional, lecture-based structure. The remaining content of the standard E\&M~I curriculum is covered at the beginning of the senior year, as a part of Electromagnetism Capstone (PH431), which also covers much of the content of a more traditional E\&M~II course.

\section{Methodology}\label{sec:method}
\subsection{The CUE diagnostic}

The CUE was originally developed as a free-response (FR) conceptual survey of electrostatics (and some magnetostatics) for the first semester of an upper-division level E\&M sequence. It is designed in a pre/post format. The 20-minute pretest contains 7 questions selected from the full post-test (17 questions) that junior-level students might reasonably be expected to  solve based on their introductory course experience. The post-test is intended to be given at the end of the first upper-division semester in a single 50-minute lecture. Instead of actually \emph{solving} problems, students are asked to \emph{choose} and \emph{defend} a problem-solving strategy. They are rated both for coming up with the appropriate method and for the correctness of their reasoning in deciding on a given method. The instructions students are presented with are as follows:
\begin{quote}
For each of the following, give a brief outline of the EASIEST method that you would use to solve the problem. Methods used in this class include but are not limited to: Direct Integration, Ampere's Law, Superposition, Gauss' Law, Method of Images, Separation of Variables, and Multipole Expansion.

DO NOT SOLVE the problem, we just want to know:\vspace{-0.5em}
\begin{itemize}
\item The general strategy (half credit)
\item Why you chose that method (half credit)
\end{itemize}
\end{quote}
The CUE contains several types of problems: ``outline method with explanation'' questions (Q1 -- Q7, Q14, Q17), ``evaluate and explain'' problem (Q8), multiple choice questions with (Q9, Q13, Q16) and without (Q15) explanation, problems requiring sketching without explanation (Q10, Q12c,d) and problems requiring only an answer without explanation (Q11, Q12a,b). Recently, the Physics Education Research (PER) group at CU has transformed the free-response version of the CUE into a multiple-choice version \cite{Wilcox13-MCC, Wilcox14-MCC}.

\begin{figure}[t]
 \includegraphics[width=0.47\textwidth]{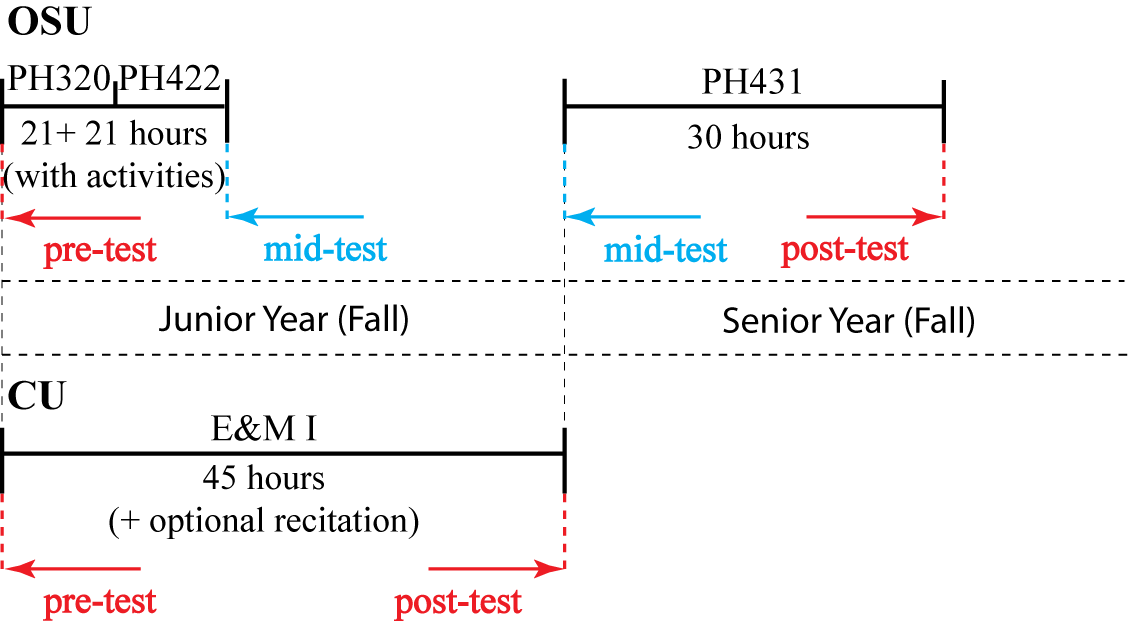}
 \caption{(Color online) Schedule of administering the CUE at OSU (quarter systems) and CU (semester system). The horizontal axis represents weeks. The CU E\&M~I course occurs over 15 weeks, whereas the PH320 and PH422 Paradigms are more intense and last 3 weeks each.}
\label{fig:emscheme}
\end{figure}

\subsection{The CUE administration}\label{subsec:CUE_admin}

For our study, we collected the CUE data over a period of four years (from 2010 to 2013). At the beginning of the Fall term of each year, junior-level students enrolled in the Symmetries and Idealizations Paradigm course (PH320) took the CUE pretest (see Fig.~\ref{fig:emscheme} for a timeline of the CUE at both OSU and CU). The same group of students was given the midtest (a subset of 12 post-test questions we chose to conform to our course goals) at the end of the Static Vector Fields Paradigm course (PH422/522). In the following year two tests were given within the Electromagnetism Capstone course (PH431). There was a second midtest at the beginning of the term (with the same set of 12 questions as in the first midtest) and the CUE post-test at the end of the term . In our analysis we followed the ``CUE rubric" v.23 \cite{CUEMI}.

The necessity of introducing the midtest arose due to the different course structure at OSU. Since not everything that the CUE tests is covered by the end of the fall quarter of the junior year, the results from a full CUE post-test would not have been appropriate.  We also note that, although OSU students have had more contact hours in E\&M (72 hours) at the time they take the post-test than CU students (45 hours), most of the additional hours are on the more advanced content (corresponding to CU's E\&M~II). We found a strong correlation between the first CUE midtest scores and final grades for PH422 ($r=0.53$, $p<0.001$, $N=85$) and no statistically significant relationship between the CUE post-test scores and final  grades in PH431 ($r=0.18$, $p>0.05$, $N=36$) \cite{Cohen-SA}. This suggests that the additional material in PH431 is not influencing students' performance on the CUE.

It has been shown that the time frame for giving a test -- i.e., administering the test \emph{at} vs. \emph{near} the beginning or the end of a course -- can have a significant effect on the test results \cite{Ding08-ECT}. With each Paradigm lasting only 3 weeks, there is not much flexibility as to when the CUE can be administered. This helps to maintain consistent testing conditions and reduce possible variations between scores when comparing data collected over multiple years. The timing of the each test was consistent throughout the whole period discussed  -- the pretest and second midtest were given during the first or second day of class and the first midtest and the post-test were given during one of the last two days of class.

\subsection{Demographics}

Over a period of four years we have administered the CUE pretest to a total of $N=100$ students, the first midtest to $N=92$ students, the second midtest to $N=91$ students and the full post-test to $N=39$. In our analysis we excluded data from two groups of students: The first were members of the PER group at OSU, who participated in meetings where the CUE diagnostic was discussed. The second were students who either withdrew during the course or took only some of OSU's E\&M courses and therefore did not take a sequence of at least two consecutive tests. This left us with $N=85$ for the pretest, $N=86$ for the first midtest, $N=69$ for the second midtest and $N=37$ for the post-test.

\begin{figure*}
 \includegraphics[width=0.75\textwidth]{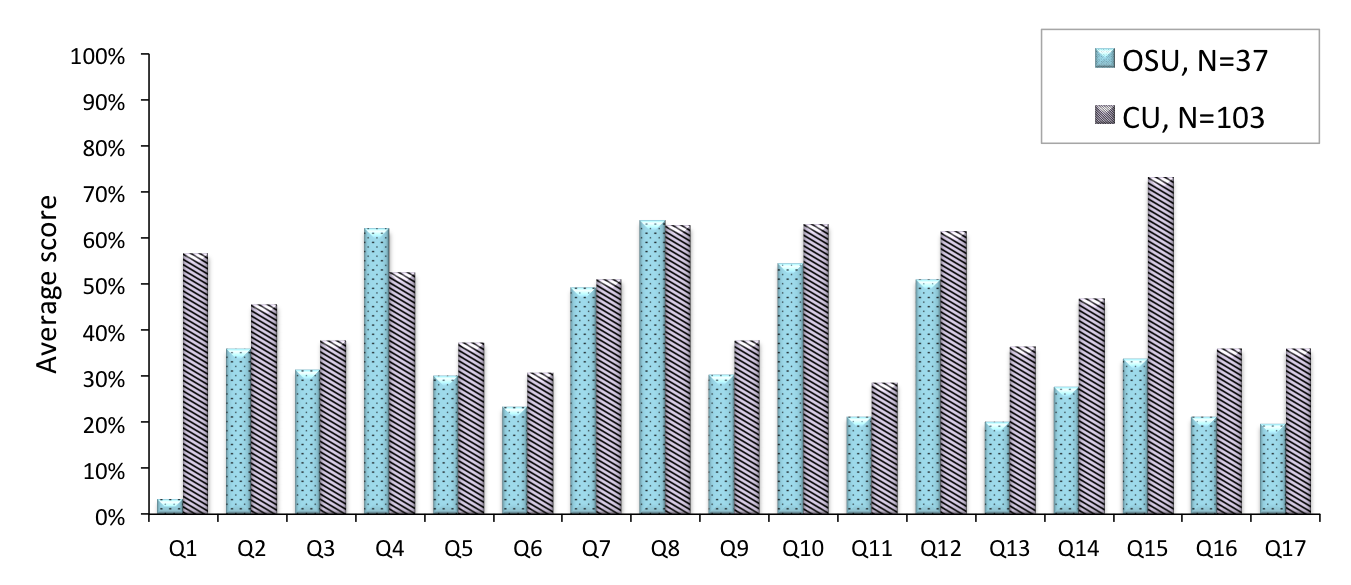}
 \caption{(Color online) Mean values for each question on the CUE post-test  for OSU ($N=37$, blue dotted pattern) and for CU ($N=103$, purple hatched pattern) students.}
\label{fig:post_total}
\end{figure*}

There were multiple instructors teaching each course -- two for PH320 (one PER and one non-PER researcher), three for PH422 (all PER researchers) and two for PH431 (both non-PER researchers). Due to its structure (intense pace, interdependence of the content between courses) there is a well defined plan to follow for the Paradigms courses. While instructors are free to introduce additional content to the course, the well-developed resources for Paradigms assure the consistency of teaching the core concepts among different instructors. In Capstones instructors have more freedom as to how the class is being taught. However, we did not find a statistically significant difference between the average CUE scores for groups with different instructors as determined by the one-way ANOVA ($F=1.08$, $p=0.34$).

\subsection{Data analysis}

For the statistical analysis we used the Statistics Toolbox of the program Matlab R2010a \cite{Matlab}. The normality of data was verified using the Szapiro-Wilk test. In order to check the difference between two sample means the paired t-test was used and for three sample means we used the one-way ANOVA. $p$ values lower than 0.05 were considered to be significant.

\section{What the CUE tells us about student learning at OSU}\label{sec:what_CUE_tells_us}

\begin{figure}[b]
\includegraphics[width=0.46\textwidth]{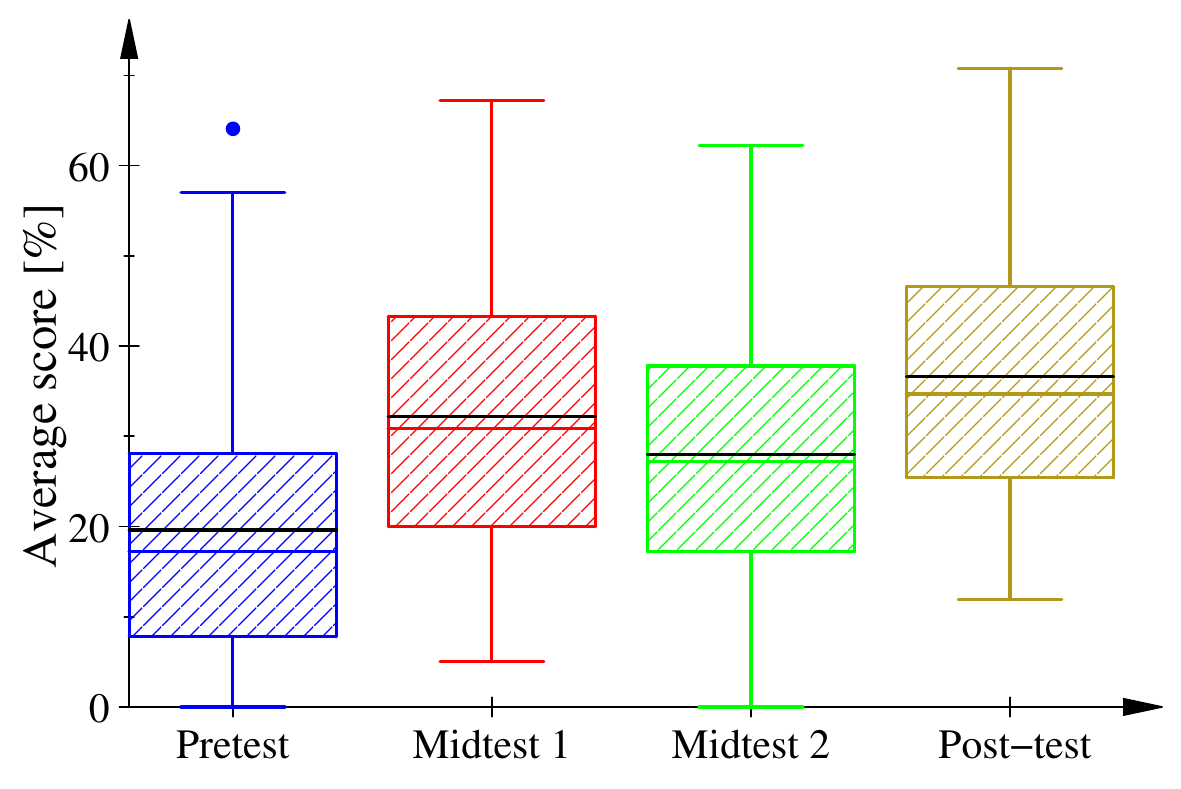}
\caption{(Color online) Box plots of the students' scores on all of the CUE tests at OSU. The pretest plot is for $N=85$; the first midtest for $N=86$; the second midtest for $N=69$; the post-test for $N=37$ students. The mean (black line) for all tests is slightly higher than the median. The central lines indicate medians for each test and the central box represents $50\%$ of the data. The lower whisker extends to either the smallest value or the $1.5$ interquartile range (IQR), whichever is greater (the IQR is calculated as a difference between the third and the first quartiles). The upper whisker extends to either the largest value or the $1.5$ IQR, whichever is smaller. For the pretest there was one unusually high score (outlier) represented as a dot at over $60\%$.}
\label{fig:box_pots_all}
\end{figure}

In this section, we discuss what the CUE results reveal about curriculum at OSU \cite{Zwolak14-RDC}. Box plots of the students' scores for all four tests are presented in Fig.~\ref{fig:box_pots_all}. One can see that, as students progress through courses relevant to E\&M, their average scores increase significantly. The drop between first and second midtests is likely due to a lack of E\&M material taught in this time frame and will be discussed further in Section~\ref{sec:gap_in_instruction}. 

Throughout this section we mainly focus on the post-test data. The CUE post-test was administered three times between the Fall term of 2010 and the Fall term of 2013 (with the exclusion of the Fall term of 2012). Figure~\ref{fig:post_total} shows a comparison of the average performance on each question between students from OSU (blue dotted plot) and CU (purple hatched plot) \cite{Note1}. One of the most striking features of this plot is the similarity of the overall pattern of students' scores -- both on the high- and low-scored questions.  With the exception of two questions (Q1 and Q15), the averages agree to within $10\%$ on the first 12 questions and to within $20\%$ thereafter \cite{Note2}. It is also worth noting that, despite the low number of students taking the CUE post-test in individual years, this pattern is still preserved when comparing the average scores on each question by year. This suggests the CUE is reliable across the two very different curricula. Moreover, the low average scores on some questions from both institutions suggest that the CUE is a very challenging test in general, regardless of the curriculum. 

\subsection{The overall results: average vs. gain}

\begin{figure}[t]
\includegraphics[width=0.37\textwidth]{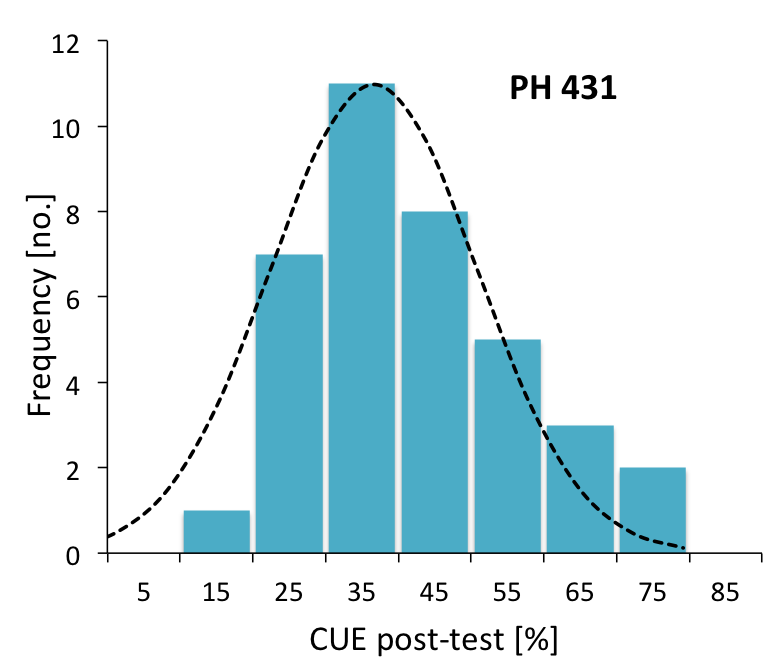}
\caption{Histogram of the students' scores on the CUE post-test based on $N=37$ students in three courses. The dotted line shows the Gaussian best fit to the data.}
\label{fig:PH431-dist}
\end{figure}

Students at OSU scored on the post-test on average $36.6\pm2.4\%$ (compared to $47.8\pm1.9\%$ at CU reported in Ref.~\cite{Chasteen12-CUE}), with the spread of their performance ranging from about $12\%$ to $70\%$. Scores are distributed normally around the mean (see Fig.~\ref{fig:PH431-dist}). The normality of the scores was verified using the Shapiro-Wilk test with $p=0.36$. 

\begin{figure}[t]
\includegraphics[width=0.47\textwidth]{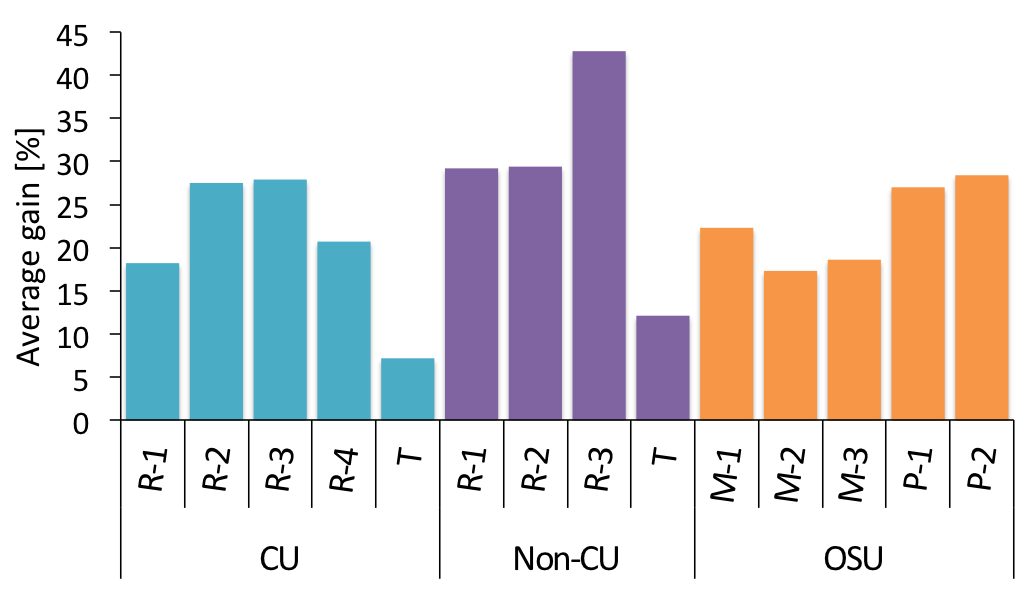}
\caption{(Color online) The average CUE gain across different institutions. R denotes the PER-based courses, T the standard lecture based courses and P the data from OSU. For comparison we present gains from the first midtest (M). Data for CU and non-CU gains adapted with permission from Ref.~\cite{Chasteen12-CUE}.}
\label{fig:gain}
\end{figure}

To provide a measure of student improvement over time we used the normalized gain proposed in Ref.~\cite{Hake97-ANG}. The non-normalized (absolute) gain is an actual average gain calculated as
\begin{equation*}
\mathbf{g}_{abs}=\langle 7Q\,post\textrm{-}test\rangle-\langle pretest\rangle\,,
\end{equation*}
where $\langle 7Q\,post\textrm{-}test\rangle$ denotes the average score of a student on the 7 post-test questions that correspond to the pretest. The normalized gain is defined as the ratio of the absolute gain to the maximum possible gain,
\begin{equation*}
\mathbf{g}_{nor}=\frac{\mathbf{g}_{abs}}{100-\langle pretest\rangle}\,.
\end{equation*}
For students who took both the pre- and post-tests ($N=24$), we found an average normalized gain of $33\%$ ($28\%$ non-normalized), which is similar to gains of $34\%$ (normalized) and $24\%$ (non-normalized) at CU reported in Ref.~\cite{Chasteen12-CUE}. The significance of this gain was confirmed using the paired t-test ($p<10^{-4}$). Thus, although students at OSU on the average scored about $12\%$ lower than students at CU on both the pre- and post-tests, they showed similar learning gains to students from other institutions taught in PER-based courses, and higher gain than observed in standard lecture-based courses (see Fig.~\ref{fig:gain})~\cite{Note3}.

\subsection{Revealing differences between curricula} 

\begin{figure}[b]
 \includegraphics[width=0.45\textwidth]{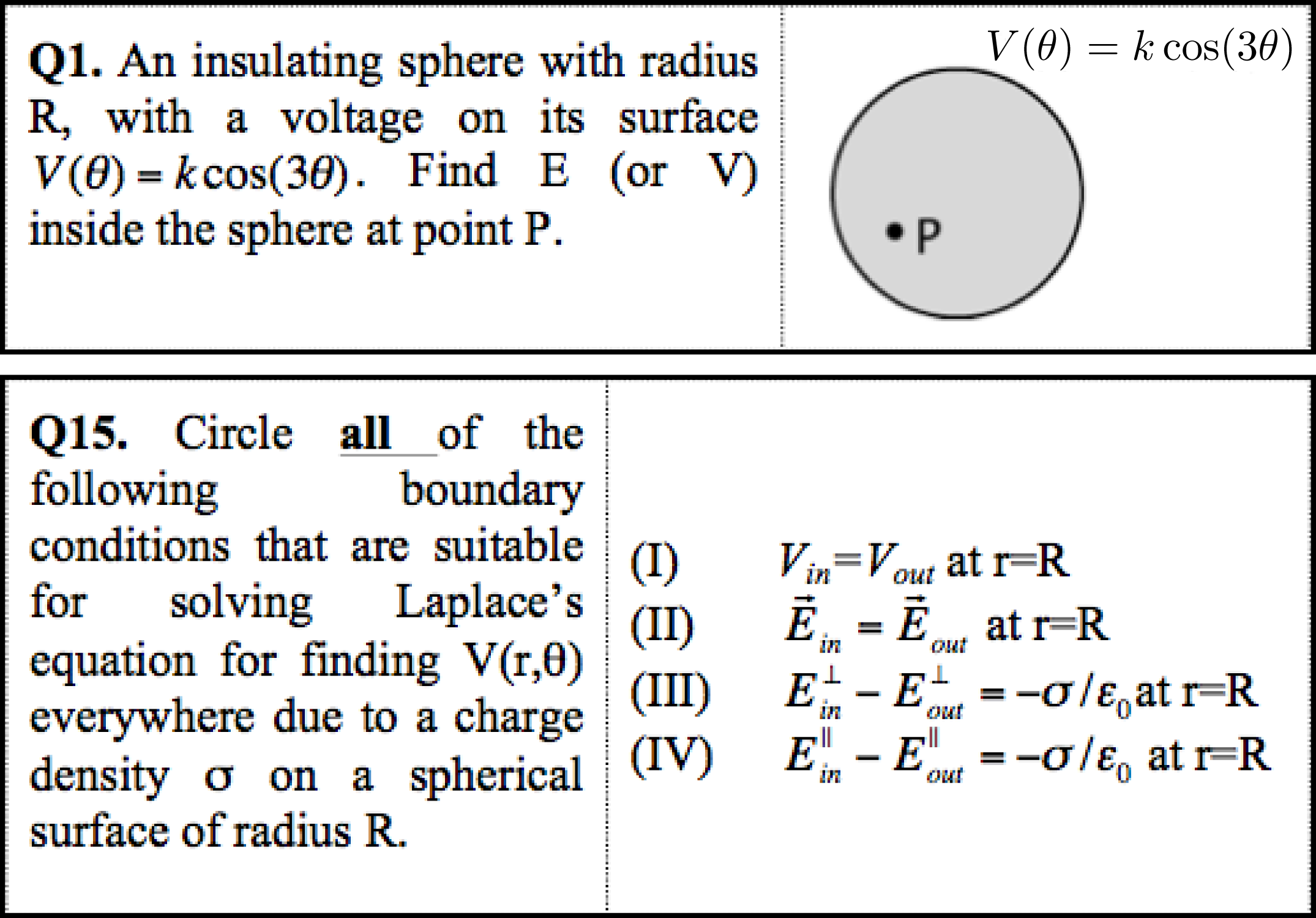} 
\caption{Questions where the scores of OSU students differ significantly from the scores of students taught at CU. Reproduced from the CUE \cite{CUEMI}.}
\label{fig:Q1andQ15}
\end{figure}

Although the overall pattern in Fig.~\ref{fig:post_total} from both institutions is very similar, there are some significant differences that need to be addressed. In particular, OSU students' scores differ by over $50\%$ on question Q1 regarding finding the potential $V$ (or field $E$) inside an insulating sphere and by almost $40\%$ on question Q15 regarding selecting boundary conditions to solve for the potential $V(r,\theta)$ on a charged spherical surface (for reference, the full problems are reproduced in Fig.~\ref{fig:Q1andQ15}). Both of these questions are intended to test whether students can set up the solution to a problem involving partial differential equations (i.e, recognizing separation of variables as an appropriate problem-solving technique and/or defining the proper boundary conditions) \cite{Chasteen12-CUE,CEMLG}. We do not find any indication that these discrepancies were due to issues with the rubric. Thus, to determine their origin, we need to look more closely at the learning goals for the relevant courses at OSU.

In a traditional curriculum, as defined by the standard E\&M text by David Griffiths \cite{Griffiths-EM}, students are often first exposed to the application of separation of variables in physics in their E\&M course, before they take quantum mechanics. At OSU, however, students are exposed to the separation of variables mainly in the context of the Schr\"{o}dinger equation -- first in the ``1-D Waves'' and the ``Central Forces'' Paradigms in the Winter term of the junior year and then in the Mathematical Methods Capstone in the Fall term of the senior year~\cite{Note3}. The separation of variables is discussed in multiple courses before students take the E\&M Capstone and thus not much time is devoted to this topic in the Capstone itself. To be precise, there is only one day (typically the second day of the first week) spent on Laplace's equation, followed by 2 or 3 homework problems (see Ref. \cite{PH431} for a detailed Syllabus for PH431). As a consequence, students have much more experience with separation of variables in the context of quantum mechanics, long before they see it as part of E\&M, and even then the structure of the Capstone does not provide them with many opportunities to practice it in the E\&M context. Low scores on the two other questions involving separation of variables and boundary conditions (BCs): Q11 (finding BCs in a specific scenario) and Q13 (recognizing the form of solutions that match given BCs) supports our suspicion that students are not getting enough exposure to these topics in the context of E\&M. Our findings agree also with previous studies that find the positive transfers of skills across context and content to be rare (see, for example, Refs. \cite{Reed74-ROT,Mestre-TOF}). 

\begin{figure}[t]
 \includegraphics[width=0.45\textwidth]{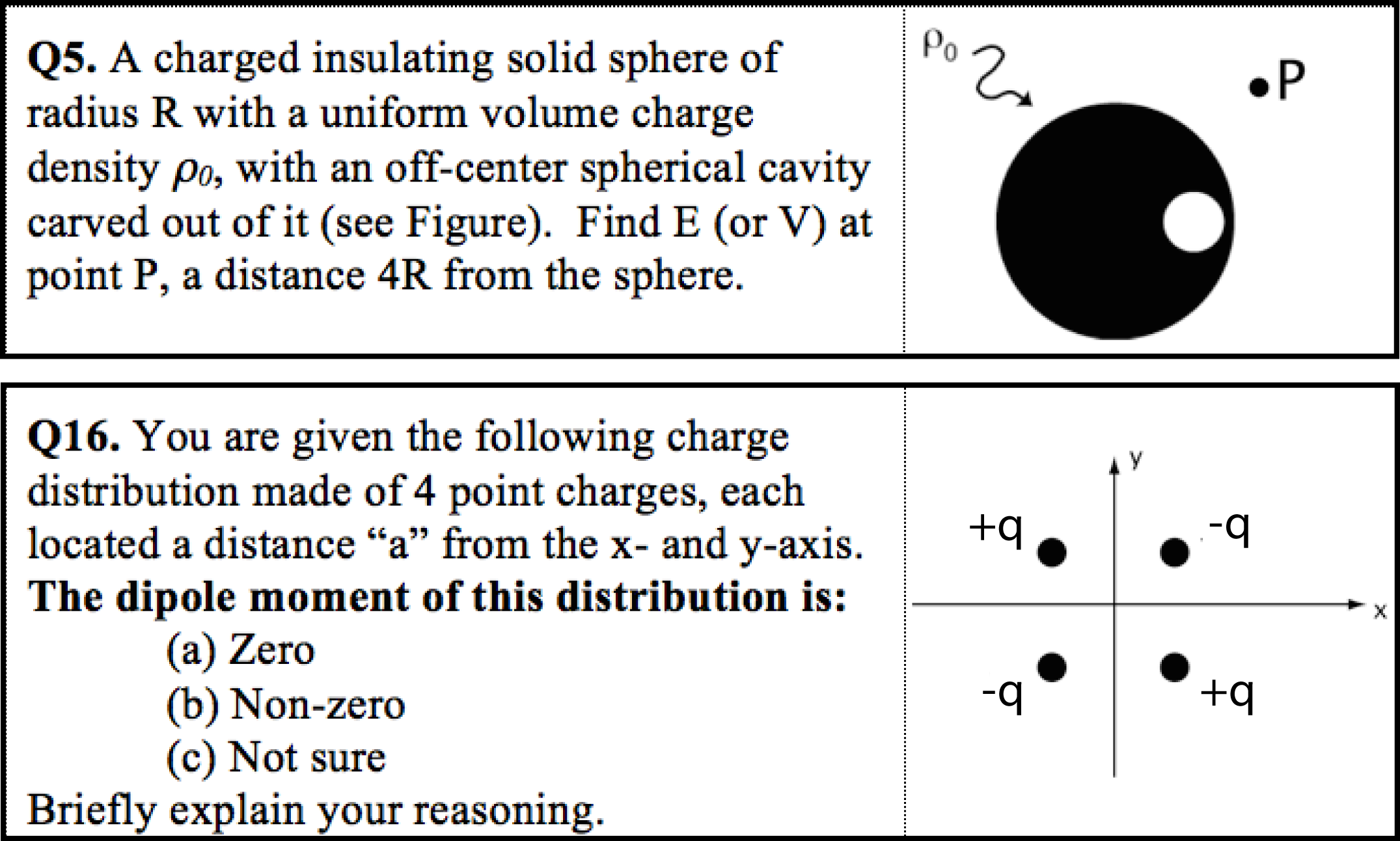} 
\caption{Questions 5 and 16 reproduced from the FR version of the CUE \cite{CUEMI}.}
\label{fig:Q5andQ16}
\end{figure}

\section{What OSU and CU data tells us about the CUE: Problems with rubric}\label{sec:what_curriculum_tells_us}

During an initial grading of OSU students, we have found that, although the questions on the CUE reflect many of our learning goals in an appropriate manner, for some questions the current rubric for the CUE is particularly aligned to the topics and methods of teaching at the University of Colorado \cite{Zwolak13-PWR}. In particular, we noticed many solutions, including ones we would view as correct, that did not seem to fit the rubric provided with the CUE. As an example we will discuss two problems from the CUE: Q5 (involving the superposition principle) and Q16 (involving finding the dipole moment of a given charge distribution). Both problems are reproduced in Fig.~\ref{fig:Q5andQ16}. 

The content related to these two questions at OSU is discussed as part of the first two Paradigm courses (PH320 and PH422). Therefore, in order to provide a reasonable comparison between CU and OSU, we looked at results from OSU on these questions given as part of the first midtest at the end of the fall term in the junior year as well as results from the full post-test.
\begin{table}[b]
\caption{Main categories of responses for our analysis. In addition to the below, we considered also ``F'' for answers that were irrelevant for our analysis, ``X'' for the lack of an answer and ``Z'' for an answer ``I don't know.''}
\begin{tabular}{p{0.25cm}p{0.5cm}p{5.5cm}}
\hline
A&\multicolumn{2}{l}{Clearly talks about adding electric fields}\\
&A1&uses the word ``superposition''\\
&A2&does not use the word ``superposition''\\
\hline
B&\multicolumn{2}{l}{Clearly talks about adding potentials}\\
&B1&uses the word ``superposition''\\
&B2&does not use the word ``superposition''\\
\hline
C&\multicolumn{2}{l}{Seems to be adding charges}\\
&C1&uses the word ``superposition''\\
&C2&does not use the word ``superposition''\\
\hline
D&\multicolumn{2}{l}{Ambiguous about what is being added/superposed}\\
&D1&uses the word ``superposition''\\
&D2&does not use the word ``superposition''\\
\hline
\end{tabular}
\label{tab:newq2/q5}
\end{table}

\subsection{The superposition principle}

Let us start with the superposition principle question (Q5). While grading the CUE tests from OSU students, we noticed that OSU students often did not use the word ``superposition,'' instead trying to explain what they would do to solve the problem. More importantly, it was often not clear from students' answers what they wanted to add/superpose -- fields, charges or something else -- even when they used the word ``superposition.'' Although the rubric accounts for situations where a student explicitly tries to superpose charges instead of fields, the ambiguous response is not accounted for in the rubric. Finally, despite the problem statement explicitly allowing for a potential approach, this approach was absent in the rubric. To address these concerns, we developed a new categorization of responses for this question, shown in Table~\ref{tab:newq2/q5}, which focuses primarily on \textit{what} is being superposed and secondarily on whether the word ``superposition'' is used.

\begin{figure}[t]
 \includegraphics[width=0.47\textwidth]{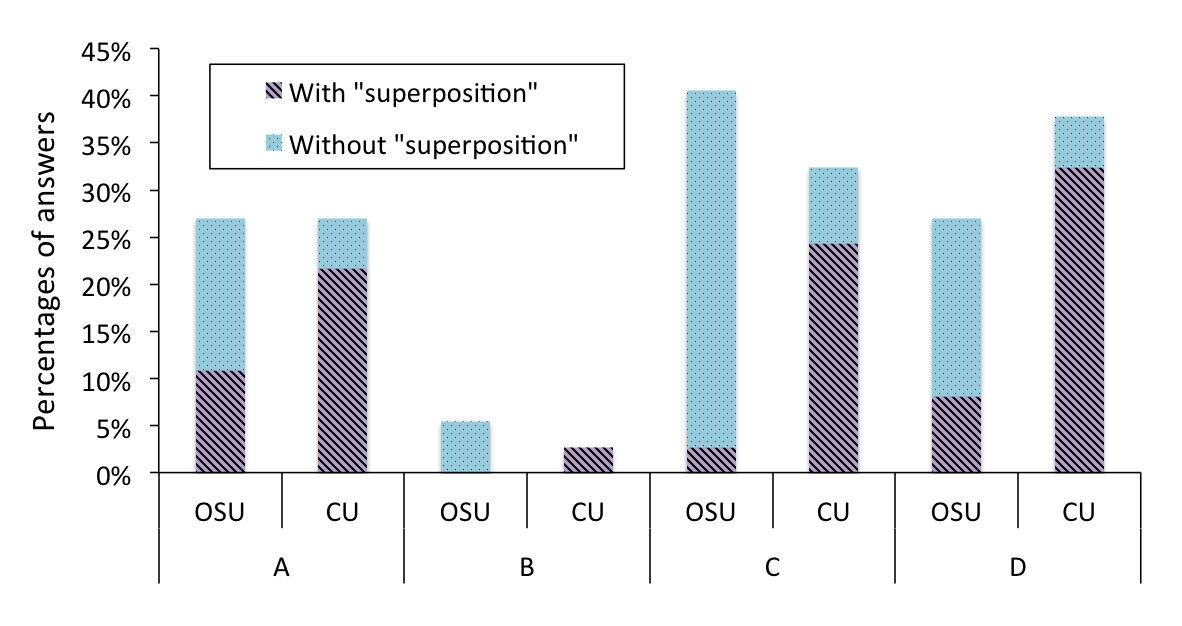}
 \caption{(Color online) Frequency of use of the term ``superposition" in the students' answers at OSU vs. CU (purple hatched pattern, $N_{OSU}=37$, $N_{CU}=37$). Explanation of categories A, B, C and D is presented in Table~\ref{tab:newq2/q5}.}
\label{fig:q5data-rel}
\end{figure}

With this new categorization, we compared responses on the superposition question for $N^{total}_{OSU}=86$ tests from OSU students and $N^{total}_{CU}=68$ tests provided by CU. In our first analysis, we considered only answers which were relevant to the problem, i.e., we eliminated the responses ``F'' (used to code answers irrelevant for the analysis), ``X'' (used to code the lack of an answer), and ``Z'' (used to code an answer ``I don't know"). This left $N_{OSU}=37$ and $N_{CU}=37$ students who tried to add/superpose something (either correctly or incorrectly). Figure~\ref{fig:q5data-rel} shows the distribution of correct answers, between the electric field approach (A) and the potential  approach (B), and incorrect answers, between clearly talking about adding charges (C) and being ambiguous about what should be superposed (D). 

The first thing to note is the difference in the explicit use of the word ``superposition.''  Of all relevant answers (combining A, B, C, D), $81\%$ of CU students explicitly used the term ``superposition,'' compared to $22\%$ students at OSU. This pattern is also evident in the correct responses (A and B only). Of all correct answers (combining A and B), $23\%$ of OSU students explicitly used the term ``superposition,'' compared to 82\% of CU students.

\begin{figure}[t]
 \includegraphics[width=0.47\textwidth]{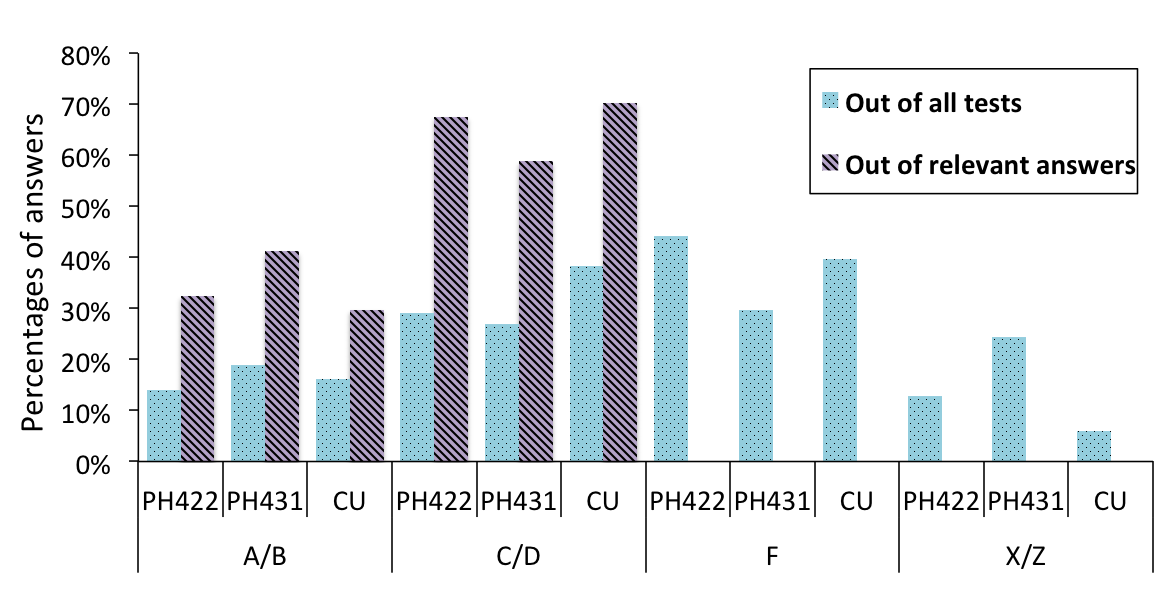}
 \caption{(Color online) Frequency of correct (A/B), incorrect (C/D), irrelevant (F) and lack of answer (X/Z) at OSU and CU out of all test ($N_{OSU}=86$, $N_{CU}=68$, blue dotted pattern) and out of only relevant answers ($N_{OSU}=37$, $N_{CU}=37$, purple hatched pattern).}
\label{fig:q5data-all}
\end{figure}

In order to look more closely at the issue of {\em what} is being superposed, we also did a comparison without considering the use of the word ``superposition'' or distinguishing between electric field and potential approaches. These results are presented in Fig.~\ref{fig:q5data-all}, which groups all correct categories (A and B) and all incorrect or ambiguous categories (C and D). The overall results are comparable for both universities. It was surprising to us that in both schools only $\sim15\%$ of all students took a clearly correct (electric or potential field) approach to this problem ($\sim 30\%$ of relevant responses). If we look only at relevant answers, in almost  $70\%$ of cases students were either unclear about what they wanted to add/superpose or were clearly talking about adding charges.

One might expect that at the institution developing the CUE there will be noticeable relationship between the test and the reformed course materials, such as clicker questions that are similar to questions on the CUE, in whole or in part. Regarding the difference in emphasizing the use of the word ``superposition'', the CU course materials, which include lecture notes, clicker questions, tutorials, \emph{etc.}, seem to strongly emphasize the term ``superposition'' \cite{CUEMI}. This emphasis is not similarly apparent in the Paradigms materials \cite{PinP}. The interaction between the development of the course materials and the development of the assessment is not unexpected, but it is important to consider when extending the assessment beyond the institution of origin.

\subsection{Free Response vs. Multiple Choice CUE: Multipole, Gauss' Law and Delta Function}

As mentioned earlier, the PER group at CU had recently developed a multiple choice (MC) version of the CUE test. The preliminary validation of this test at CU showed that for most questions (all but four) there are no statistically significant differences between the FR and MC versions at CU \cite{Wilcox13-MCC}. 

\begin{figure}[t]
 \includegraphics[width=0.47\textwidth]{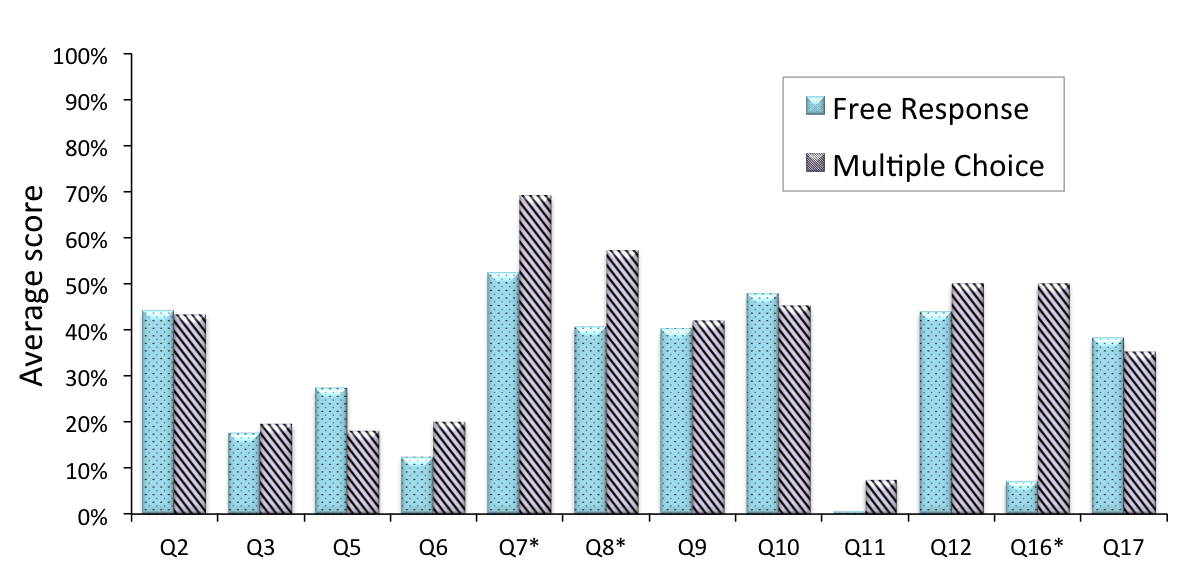}
 \caption{(Color online) Comparison of OSU students average scores between FR ($N=86$, blue dotted pattern) and MC ($N=21$, purple hatched pattern) versions of the CUE. There are significant differences in scores on questions Q7 (Gauss' Law), Q8 (Delta function) and Q16 (dipole moment, Q15 on MC), marked with an asterisk.}
\label{fig:FRvsMC}
\end{figure}

The pretest version of this test was administered at OSU in a Fall term of 2013 to $N=30$ students and the midtest version to $N=21$ students. Comparison of average scores for both versions of the midtest are presented in Fig.~\ref{fig:FRvsMC}. We found significant differences in scores for three questions. On the question regarding Gauss' law (Q7) students scored on average $52.5\%$ (FR) vs. $69\%$ (MC). On the question regarding the Delta function (Q8) they scored $40.7\%$ (FR) vs. $57.1\%$ (MC). The biggest difference was on the question regarding the dipole moment (Q16 on FR, Q15 on MC), where we observed 7-fold increase in the average score on the MC version of the CUE. We note that the CU reported discrepancies on different questions (for details, see Ref.~\cite{Wilcox13-MCC}).

\begin{figure}[t]
 \includegraphics[width=0.49\textwidth]{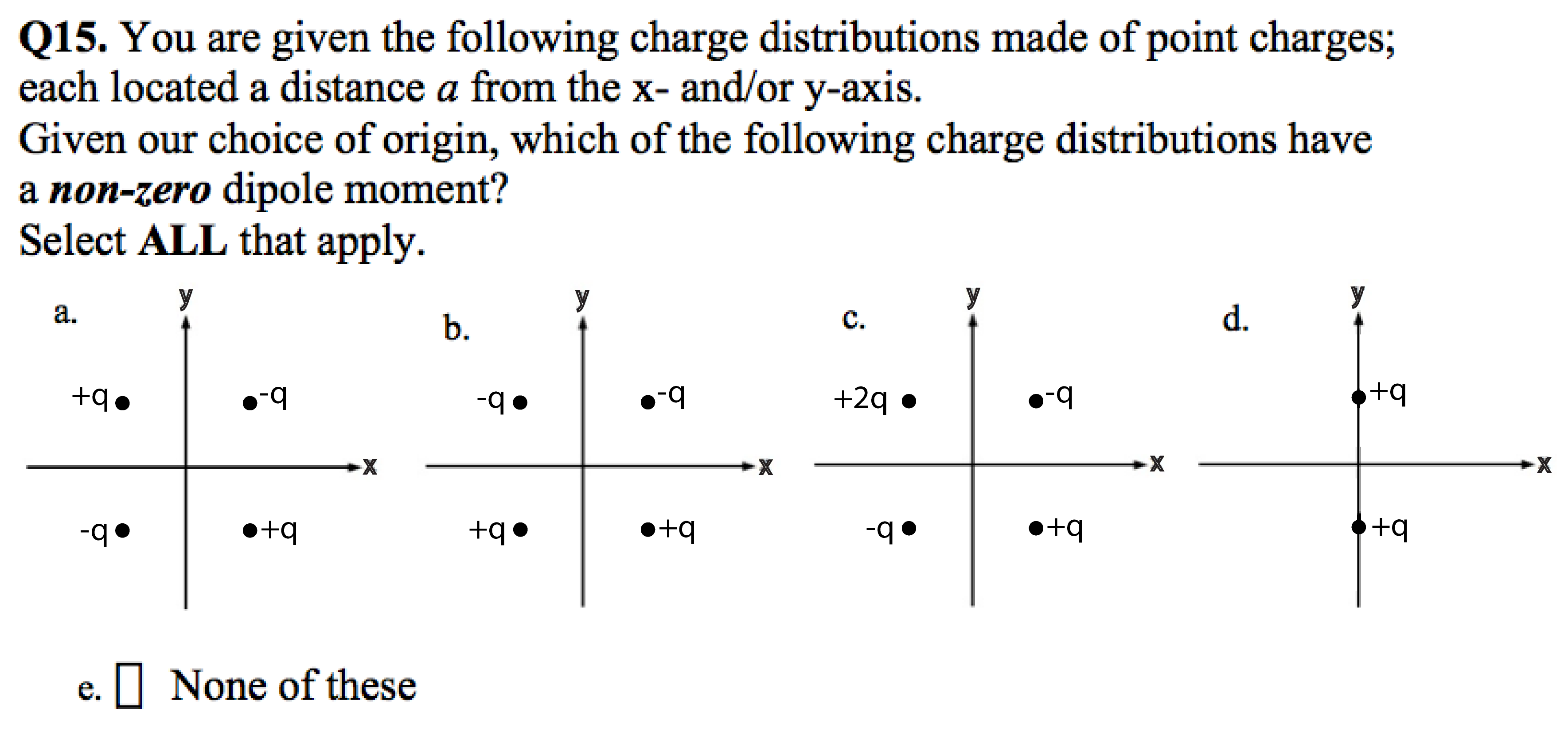}
 \caption{Question 15 reproduced from the MC version of the CUE \cite{CUEMI}.}
\label{fig:Q15MC}
\end{figure}

We start our discussion with the dipole moment problem (Q16 on the FR version, Q15 on the MC version of the CUE). On the first FR midtest OSU students scored on this problem on average $7.2\pm1.9\%$ ($N=86$). On the MC version the average score on this question increased to $50\pm9.4\%$. The objective of this problem changed in the MC version of the CUE from deciding whether the dipole moment in a particular distribution is zero to deciding which of the four presented distributions has a vanishing dipole moment (see Fig.~\ref{fig:Q5andQ16} and Fig.~\ref{fig:Q15MC}). While this change did not lead to inconsistency in scores between FR and MC versions at CU (increase of $2-3\%$ on MC version), average scores at OSU changed significantly. OSU students' MC midtest score is actually higher even than the FR post-test. 

The reason for this discrepancy remains an open question. One possible explanation for such a big difference is that on the FR version many OSU's students did not attempt to solve this problem at all  ($17.4\%$) or gave the ``I don't know" answer ($24.4\%$). In the MC version only $9.5\%$ of students left this question unanswered. This result is consistent with the idea that it is easier to recognize an answer than to generate it \cite{Semb93-LTM,Cabeza97-RAR}. 

Another possible reason for significantly lower scores on the FR version is that some OSU students used the symmetry of the system, without further explanation, as an argument for choosing a vanishing dipole moment  ($17.1\%$). Since the rubric for the full answer requires mentioning oppositely directed dipoles for which the sum of dipole moments gives zero, the ``symmetry" answer is insufficient. Prior to the first midtest, OSU students take the ``Symmetries'' Paradigm, in which emphasis is placed on using symmetry arguments in various scenarios and therefore for those students the ``symmetry" argument may seem sufficient to support their choice. The significantly higher score on the MC version shows that, when presented with multiple charge distribution scenarios, students indeed recognize the ones with an appropriate arrangement of charges.

We observe a similar situation in the case of the other two questions. On both these questions, the students' average score on the MC CUE was over $16\%$ higher than on the FR CUE. If we look separately at the answer and the explanation scores for these questions, we can see a big discrepancy between scores for each part. While students scored on average $52.5\%$ for Q7, they scored $70.2\%$ for recognizing Gauss' Law as the correct method but only $27\%$ for the explanation. On Q8 students averaged $49.3\%$ for  correctly integrating the delta function but only $27\%$ for recognizing the correct physical situation. The high score on the MC CUE shows that, once presented with a set of answers with distractors, OSU students can correctly identify the right reasoning but it is much more difficult for them to come up with a reasoning that fits the rubric. We discuss this issue further in Section~\ref{sec:summary}.

\section{What the midtest tells us about learning gains during a gap in instruction}\label{sec:gap_in_instruction}

While pre- and post-testing is currently a standard approach to assess student learning gains, it fails to reveal the dynamics of student learning. One way to better understand the evolution of students learning is to repeatedly measure student comprehension of the content throughout the course and to compare it to what is actually taught in the course at a specific time. Recently this approach has been used in research on the decay of student knowledge in introductory physics courses \cite{Sayre09-PDK,Heckler10-MMU,Sayre12-LRF}. Testing only at the beginning and at the end of a course also does not reveal the changes in student performance beyond the duration of the course. The time dependence of learning is subtle and even significant gains are sometimes short lived \cite{Postman73-CII,Semb93-LTM,Bouton93-CTM}. 
 
While the intense pace of the Paradigms makes it difficult to collect data from surveys throughout the course, the unique course structure at OSU gave us the opportunity to introduce an additional CUE test -- the midtest version of the CUE discussed in Section~\ref{subsec:CUE_admin}.

Since the two midtests are administered within 10 months from each other, they provide insight into how much students forget (or learn) over the period between the end of the Fall term of their junior year and the beginning of the Fall term of their senior year, when they are not formally enrolled in any E\&M-related course but are quite intensely studying physics. It also allows us to look at long-term learning in the Paradigms curriculum. 

\begin{figure}[t]
 \includegraphics[width=0.47\textwidth]{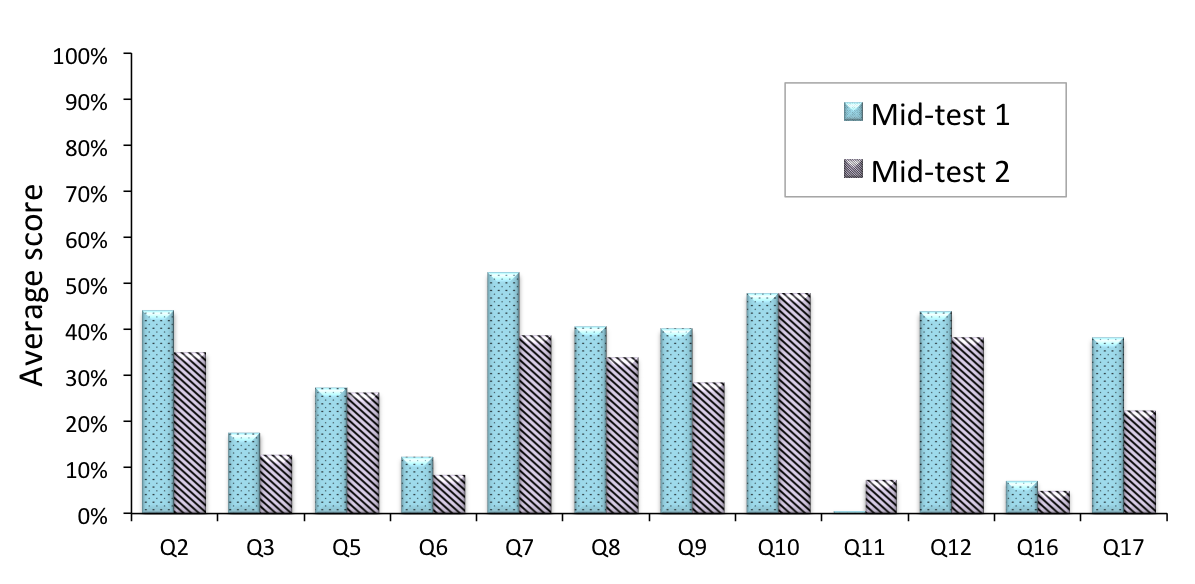}
 \caption{Comparison of students average scores between the first (blue dotted pattern) and second (purple hatched pattern) CUE midtests ($N=57$).}
\label{fig:mid1-vs-mid2}
\end{figure}
 
Figure~\ref{fig:mid1-vs-mid2} presents average scores for $N=57$ students who took both midtests. When compared to the first midtest, students lost on average $9.7\pm2.4\%$, wherein $N=18$ improved their scores by $10\%$ on average and $N=39$ had scores lower than on the first midtest by $19\%$ on average. To adjust for their initial learning, one can look at the relative percentage loss, $\ell_{rel}$, defined as
\begin{equation*}
\ell_{rel}=\frac{\langle midtest\,2\rangle-\langle midtest\,1\rangle}{\langle midtest 1\,\rangle}\cdot100\%\,,
\end{equation*}
where $\langle midtest\,1(2)\rangle$ denotes the average score of a given student from a full midtest 1(2). Students at OSU showed an average loss of $17.3\pm4.2\%$. This data is consistent with previous research on long-term learning, showing that students retain approximately $85\%$ of what they had learned after 4 months and about $80\%$ after 11 months.

While the observed forgetting rate is not unusual, one can ask if there are other factors that lower the second midterm average. The timeline of administering the CUE at OSU  (Fig.~\ref{fig:emscheme}) shows that the first midtest is administrated in a different circumstances than the second one. Students take the first midtest in the middle of the quarter, when they are still learning and they might be trying to do their best and to solve as many problems as they can. The second midtest, on the contrary, is administered at the beginning of the Fall term in senior year, right after the summer break. Based on the number of ``I don't know" (code ``Z'') and blank (code ``X'') answers, students seem to be taking this test more casually and do not try to answer when they are not sure. The proportion of ``X'' and ``Z'' answers on the second midtest reaches between $20\%$ and $46\%$ for six out of 12 questions while on first midtest all question but one (Q11) have ``X'' and ``Z'' percentage rate of less than $15\%$. Moreover, $95\%$ of students taking the first midtest declared they took it ``seriously'' or `` somewhat seriously'' compared to $78\%$ of students taking second midtest. The lower scores on the second CUE midtest thus might be a reflection of forgotten knowledge in combination with other factors, such as a more informal atmosphere.

\section{Summary}\label{sec:summary}

The Colorado Upper-Division Electrostatics diagnostic is meant to serve as a tool to assess student conceptual learning in E\&M at the junior level. It has been validated in multiple institutions, in both PER and non-PER based courses, providing reliable and valid information about the achievement of students under junior-level E\&M instruction \cite{Chasteen12-CUE}. 

Due to the significantly restructured curriculum at OSU, our findings provide valuable data for comparison with results from CU's more moderately reformed curriculum and from institutions with a more traditional (lecture) format. While the sample of students at OSU is quite different from CU's students in terms of the program of study and the teaching methodology, the difficulty pattern, shown in Fig.~\ref{fig:post_total}, for most questions is preserved. This result confirms the overall robustness of the CUE. In addition, the strong differences in scores on a few specific questions shows that this assessment test is also capable of helping to distinguish between different programs of study and uncovering important gaps in a curriculum. The CUE not only recognizes what problems students are struggling with, but also sheds light on how the performance of students under reformed curricula, such as  Paradigm in Physics, compares to the performance of students taught in more traditional courses. 

It is crucial to understand the causes for the large differences between scores on particular questions. As we indicated above, one of the reasons for such discrepancies on Q1 and Q15 in the case of OSU might be the current organization of courses. While restructuring the junior- and senior-level program of study at OSU, it was assumed that -- once exposed to certain techniques of solving problems in one context -- students will be able to transfer their knowledge of its applicability from one subfield of physics to another. As the CUE has revealed, however, this is not happening and the separation of variables procedure does not become a natural E\&M problem-solving technique for students when they depart from the quantum world. To address this issue, OSU has made a recent change in the schedule of the Paradigms and Capstones -- moving the Mathematical Methods Capstone, as well as the ``Central Forces'' Paradigm to the Spring term of the junior year. This rearrangement gives us an opportunity to test whether the inclusion of more examples where the separation of variables and boundary conditions are explicitly used to solve E\&M problems can impart the generality of the techniques to the students and subsequently be reflected in higher CUE scores on the relevant questions. We are currently collecting data on how this change affects the students' performance and  will discuss this in a later publication. 

Due to the open-ended form of the original version of the CUE, its grading is a quite challenging and time-consuming task.  As we pointed out earlier, in its current form the rubric has flaws that make it difficult to consistently grade some of the questions. Moreover, while the FR CUE is designed to test whether students can generate particular arguments rather than recognize them, we showed that students taught in accordance with different curricula might present their reasoning in a form that will not fit the rubric, indicating a need to revise the rubric on some questions. The MC version of the CUE helps with the former problem as it is easy to be consistent with grading a multiple choice test. The preliminary analysis of the MC data shows significant improvements on questions that were particularly difficult to grade on the FR version (e.g., Q7, Q15). It is easier for students to decide on the appropriate answer/reasoning rather than generate one that will use the required vocabulary or justification, both of which are highly dependent on the particular teaching approach and instructor. The MC test has its own drawbacks, such as a limited number of options to choose from (there is typically more than one method to solve a problem and not all of them can be captured within a fixed number of choices). While we have only analyzed data from the first midtest of the MC CUE, we are planning to continue data collection with this version of the CUE diagnostic to compare it with the FR CUE data from OSU and other institutions. We want to look into how changing the course schedule and the format of the assessment will affect the students' scores.

The students' understanding of particular content is dynamic and time dependent \cite{Sayre09-PDK}, both on the short \cite{Sayre12-LRF} and on the long-time scales \cite{Pollock09-LTL}. We showed how a CUE midtest can be used to track long-term, inter-instructional student learning and to assess student learning beyond the duration of the course. In particular, the midtest data allowed us to demonstrate that our students retain over $80\%$ of what they had initially learned after not having any E\&M-related courses for about 10 months. Such an analysis was possible due to the transformation of the curriculum at OSU, where E\&M content in not taught in two consecutive courses. 

The CUE diagnostic is valuable in assessing student learning and determining differences and gaps in curricula. Its 17 questions on E\&M content can be examined both independently and together to investigate different aspects of learning and teaching. The results of some questions at OSU have pointed out strengths and shortcomings in the curriculum, whereas the results of other questions have pointed to potential issues with the rubric. This knowledge can be used to improve programs of study and the students' learning outcomes. This new measure, however, is still in need of fine tuning so that it can be used universally to diagnose student progress and performance.

\begin{acknowledgments}
Supported in part by NSF DUE 1023120 and 1323800. We would like to thank Steve Pollock and Bethany Wilcox for conversations about the design and grading of the CUE and Stephanie Chasteen for helping us with CU test data. Many thanks to Anita Dabrowska who provided feedback and recommendations on statistical analyses in this paper. We also thank the anonymous referees for valuable comments.
\end{acknowledgments}

%

\end{document}